\documentclass[12pt,preprint]{aastex}
\usepackage{epsfig}
\usepackage{rotating}
\begin{document}
\title{
Ultraviolet Emission from the Millisecond Pulsar
J0437$-$4715\footnotemark[1]
\footnotetext[1]{Based on observations made with the NASA/ESA Hubble
                 Space Telescope, obtained at the
                 Space Telescope Science Institute, which is operated by
                 the Association of Universities for Research in Astronomy,
                 Inc., under NASA contract NAS 5-26555. These
                 observations are associated with program \# GO-9098.}}

\author{Oleg Kargaltsev\altaffilmark{2}\altaffiltext{2}{
Dept.\ of Astronomy and Astrophysics, The Pennsylvania State
University, 525 Davey Lab, University Park, PA 16802;
green@astro.psu.edu, pavlov@astro.psu.edu}, George G.\
Pavlov\altaffilmark{2},
and Roger W.\
Romani\altaffilmark{3}\altaffiltext{3}{Stanford University, Dept.\ of Physics,
Stanford, CA 94305; rwr@astro.stanford.edu}
}

\begin{abstract}
We observed PSR J0437$-$4715 with the FUV-MAMA detector
of the Hubble Space Telescope Imaging
Spectrometer (STIS)
to measure the pulsar's spectrum and pulsations.
For the first time, UV emission from a millisecond pulsar is
detected. The measured flux,
$(2.0\pm 0.2)\times 10^{-15}$ erg s$^{-1}$ cm$^{-2}$
in the 1150--1700 \AA\ range, corresponds to the luminosity
$L_{\rm FUV}=(4.7\pm 0.5)\times 10^{27}$ erg s$^{-1}$, for the
distance of 140 pc and negligible interstellar extinction. The shape of the
observed spectrum suggests thermal emission from the neutron star
surface with a surprisingly high temperature of about $1\times 10^5$ K,
above the upper limit on the surface temperature of the younger 
``ordinary'' pulsar J0108$-$1431. For
the few-Gyr-old J0437$-$4715, such a temperature requires
a heating mechanism to operate. The spectrum of J0437$-$4715 shows marginal evidence
of an emission line at 1372 \AA, which might be a gravitationally redshifted
 Zeeman component of the Hydrogen Ly$\alpha$ line in a magnetic field
$\sim 7\times 10^8$ G.
No pulsations
are detected, with a $3\sigma$ upper limit of 50\% on pulsed fraction.
\end{abstract}
\section{Introduction}
Millisecond (recycled) pulsars are very old
neutron
stars (NSs) spun up by accretion in binary systems. So far, X-ray
observations have been the only source of information about
emission from millisecond pulsars (MSPs) outside the radio band
(Becker \& Pavlov 2001; Becker \& Aschenbach
2002). Based on their X-ray pulse profiles and spectra, the
X-ray emitting MSPs
can be divided into two distinct groups. The pulsars from the
first group (e.g., PSR B1821$-$24, B1937+21, J0218+4232) show
X-ray pulse profiles with narrow peaks, resembling those seen in
radio,
 and large pulsed fractions, $\gtrsim 50\%$.
They have hard power-law
spectra, with photon indices $\Gamma
= 1$--2, and high spin-down luminosities, $\dot{E}\sim
10^{35}$--$10^{36}$ erg s$^{-1}$. The X-ray radiation from these
MSPs is interpreted as nonthermal emission produced in the pulsar
magnetosphere.

The second group consists of MSPs with smoother X-ray pulsations,
lower pulsed fractions, and smaller $\dot{E}$ ($\sim
10^{33}$--$10^{34}$ erg s$^{-1}$). In those few cases when the
X-ray spectra are available (e.g., PSR J0437$-$4715 and
J0030+0451), they cannot be fitted with a single power-law model. The
fits with simple spectral models (power-law, blackbody)
require at least two spectral components, one of which
is very soft. Likely, this soft component can be interpreted as
thermal emission from NS polar caps, with a temperature $\sim 1$
MK. Such polar caps, heated by a backward flow of relativistic
particles accelerated in the magnetosphere above the NS magnetic
poles,
are predicted by virtually
all pair-cascade pulsar models (e.g., Ruderman \& Sutherland 1975;
Arons 1981; Harding \& Muslimov 2002). The thermal component
cannot be seen in the first group of MSPs because it is buried
under the stronger nonthermal component, similar to young ordinary
pulsars (e.g., Pavlov, Zavlin, \& Sanwal 2002).

Detailed X-ray studies of PSR J0437$-$4715, the brightest MSP of
the second group, suggest that the thermal component of its
radiation is emitted from a region with a nonuniform temperature,
decreasing from the
magnetic poles towards the equator
(Zavlin \& Pavlov 1998; Zavlin et al.\ 2002).
Such a nonuniformity could be interpreted as due to
a heat flow 
away from the polar cap.
In addition to the
external (polar-cap) heating, a variety of internal heating mechanisms
can operate in the NS interiors, such as
dissipation of the NS rotational energy and
magnetic field
(Schaab et al.\ 1999, and references therein). Consequently, the
temperature distribution and, particularly, 
the lower value of the surface temperature
depend on the thermal conductivity of the NS matter and the
relative contributions from the external and internal heating. X-ray
observations mainly probe thermal emission from the hot polar
regions, being less sensitive to the emission from the rest of the
NS surface with a lower temperature. This low-temperature emission
can only be observed in the optical-UV range. Measuring the
temperature of the NS surface in MSPs is important because it can
constrain the NS heating models and provide information about the
physical processes operating in the NS interior. If the
magnetospheric component dominates in the optical-UV,  its
detection would help elucidate the properties of relativistic
particles in the MSP magnetospheres.

Most MSPs reside in binary systems, usually with a
low-mass white dwarf (WD) companion that, as a rule, is expected
to be brighter in the optical than the MSP itself. Therefore,
solitary MSPs look more suitable for studying the NS optical
emission. However, no firm detections of optical/UV emission from
solitary MSPs have been reported. Even very deep VLT observations
of the solitary MSPs J0030+0451 (Koptsevich et al.\ 2003) and
J2124$-$3358 (Mignani \& Becker 2003) gave negative results,
putting some constraints on the nonthermal emission in the
optical. On the other hand, if the companion of a binary MSP is
sufficiently cold, it is very faint in the UV range,
which can be used to observe NSs in nearby binary MSPs,
particularly their thermal emission.

The best target for such observations is
PSR J0437--4715, the nearest and the brightest binary MSP
($P=5.76$ ms, $d=139\pm 3$ pc,
$\tau\equiv P/(2\dot{P}) = 4.9$ Gyr,
$\dot{E}\equiv 4\pi^2I\dot{P}P^{-3}=3.8\times 10^{33} I_{45}$ erg s$^{-1}$
 --- van Straten et al.\ 2001).
Its binary companion is a cold WD with the effective surface
temperature of about 4000 K (Danziger et al.\ 1993) and orbital
period of 5.5 days. Optical emission from the binary is dominated
by the WD ($R= 20.1$, $V=20.8$, $B= 22.2$), making optical
detection of the pulsar impossible. This prompted us to carry out
observations of the system in the far-ultraviolet (FUV) range with
the Hubble Space Telescope ({\sl HST}). In this paper we report
first detection of UV emission from a non-accreting MSP.
The details of the observations and the data
analysis are presented in \S2 and \S3. The results and their
implications are discussed in \S4 and summarized in \S5.
\section{Observations}
PSR J0437$-$4715 (J0437 hereafter)
was observed on 2001 August 24 (start date
52\,145.23638667 MJD) with the Space Telescope Imaging
Spectrograph (STIS). The source was imaged on the Far-Ultraviolet
 Multi Anode Micro-channel Array (FUV-MAMA). The
low-resolution
 grating G140L (which covers the wavelength interval $\approx1150$--1700 \AA)
with the $52'' \times 0\farcs5$ slit were used. The data were taken during
five consecutive orbits, including target acquisition.
We used the WD companion of the pulsar as the acquisition
target, centering it in the slit to a $\pm 0\farcs01$ nominal accuracy. The total
scientific exposure time was 14,150 s (2,150 s for the first
exposure and 3,000 s for each of the subsequent exposures).
FUV-MAMA was operated in TIME-TAG mode which allows the photon
arrival times to be recorded with 125 $\mu$s resolution.
\section{Data analysis}
For each exposure, we processed the raw, ``high-resolution'' images
($2048\times2048$ pixels; plate scale of $0\farcs0122$ per pixel
--- see \S11 of the STIS Instrument Handbook
\footnote[4]{http://www.stsci.edu/hst/stis/documents/handbooks/currentIHB/c11\_datataking2.html
}, IHB hereafter)
using the calibration files available on January 23, 2003.
As an output, we obtained flat-fielded
``low-resolution'' ($1024\times1024$ pixels; plate scale
$0\farcs0244$ pixel$^{-1}$; spectral resolution 0.58 \AA\
pixel$^{-1}$) images and used them for the
spectral analysis.

The processed
images show a nonuniform
detector background that
consists of a flat (constant) component and the so-called
``thermal glow'' component (Landsman 1998) that
dominates over most of the detector area.
The thermal glow is the strongest in the upper-left
quadrant of the detector (Fig.\ 1), where the dark count rate can
exceed the nominal value,
$6\times10^{-6}$ counts s$^{-1}$ pixel$^{-1}$, by a factor of 20.
The brightness of the thermal glow increases exponentially with
increasing the FUV-MAMA power supply temperature above 38.9 K
(Landsman 1998). In our observation, this temperature was rising
from 40.89 K during the first orbit to 42.36 K during the last
orbit, resulting in brightening thermal glow. In addition, the
overall shape and small-scale structure of the thermal glow varied
slightly between the exposures.

Because of the high background, the dispersed source spectrum is
hardly discernible in the separate exposures. Nevertheless, we find
the spectrum centered at $Y=351\pm 2$ pixels in each of the
flat-fielded images (the centroid slightly varies with $X$), where
$X$ and $Y$ are the image coordinates along the dispersion and
spatial axes, respectively. This is within 2 pixels (0\farcs05) of
the nominal position where the target was expected to be acquired.
At this location on the detector, the contribution of the thermal
glow to the background is still quite large --- the background
averaged along the dispersion direction exceeds the nominal
background by a factor of 10. To improve the signal-to-noise ratio (S/N),
we combined the images from five exposures into a single
image using the STSDAS\footnote[5]{Space Telescope Data Analysis
System available at http://stsdas.stsci.edu/STSDAS.html}
 task ``mscombine''.
The $Y$-positions of the centroids differ by less than 3 pixels
for different exposures and different wavelengths ($X$-positions).
The spectrum of the source is clearly seen in the summed image
shown in Figure 1.

Accurate subtraction of the enhanced, nonuniform background is
crucial to measure the spectrum of our faint target. The spectral
extraction algorithm implemented in the standard STIS pipeline
(task X1D) does not adequately correct for the nonuniform
background while extracting the spectrum of such a faint source.
Therefore, we developed an IDL routine
with additional capabilities of grouping and fitting the
background to reduce the data.

Since the source spectrum occupies only a small region on the
detector, we do not attempt to subtract the background globally.
Instead, we scan the count distribution within two strips,
$324\leq Y\leq 343$ and $362\leq Y \leq 381$, adjacent to the
source region, $344\leq Y\leq 361$. To obtain the spectrum with a
sufficiently high S/N, we have to bin the spectrum heavily;
after some experimenting, we chose four spectral bins
($\lambda$-bins): 1155--1187, 1248--1283, 1316--1376, and
1400--1702 \AA. The
first two $\lambda$-bins
are chosen to avoid contamination by  the geocoronal line emission
(Ly${\alpha}$ line at 1216 \AA\ and the Oxygen line at 1304
\AA; the other geocoronal Oxygen line, at 1356 \AA, is not seen in this
observation).
Because of an enhanced background at
1376--1400 \AA, we also exclude this region from the spectral
analysis, which determines the choice of the third $\lambda$-bin
(1316--1376 \AA). The remaining counts are grouped into a single
bin (1400--1702 \AA) to have comparable S/N
in the second through fourth $\lambda$-bins.

For each of the $\lambda$-bins, we calculate the total number of
counts, $N_t$, within the extraction boxes of different heights
(one-dimensional apertures): $A_{s}=3$, 5, 7, 9, 11, 13, 15, and 17
pixels, centered at $Y=351$, 352, 353, and 352 for the 1st,
2nd, 3rd, and 4th $\lambda$-bins, respectively. To evaluate the
background, we first clean the background strips (see above) from
outstanding ($>10^{-3}$ cts s$^{-1}$ pixel$^{-1}$) values (``bad
pixels'') by setting them to local average values (median
filtering or standard sigma-clipping algorithms are not applicable
in this case --- see IHB \S7.4.2). Then, for each of the
$\lambda$-bins, we fit the $Y$-distribution of the background
counts with a first-order polynomial (interpolating across the
source region), estimate the number $N_b$ of background counts
within the source extraction aperture $A_s$, and evaluate the
number of source counts, $N_{s}=N_t-N_b$
(Table 1).

The uncertainty $\delta N_s$ of the source counts can be evaluated
as $\delta N_s = [N_t + (\delta N_b)^2]^{1/2}$, where $\delta N_b$
is the background uncertainty
in the source aperture. For a
uniform, Poissonian background, this quantity can be estimated as
$\delta N_b^{\rm uni} = [N_b (A_s/A_b)]^{1/2}$, where $A_b$ ($=40$
pixels) is the aperture where the background was measured. To
account for the background nouniformity, we estimated $\delta N_b$
directly from the image.
We binned the
distribution of background counts along the $Y$-axis with the bin
sizes equal to $A_s$ (for $A_{s}=3$, 5, and 7) and calculated
$\delta N_b$ as root-mean-square of the differences between the
actual numbers of background counts in the bins and those obtained
from the fit to the background.
The background uncertainties obtained in this way are
systematically larger (see Table 1) than the uncertainties
estimated assuming a uniform, Poissonian 
background.

We find that the signal-to-noise ratio, $S/N =
N_{\rm{s}}/\delta N_{\rm{s}}$, calculated
for each $\lambda$-bin and for
different extraction box heights, is the largest  for
$A_s=3$ or 5 pixels, depending on the $\lambda$-bin. Since the results at
small heights are sensitive to the deviation of the spectrum from
the straight line (along the dispersion direction) and possible
small misalignments between the frames taken in the different
orbits, we choose $A_s=5$ pixels as the optimal value.
We adopt this source extraction aperture for further analysis,
which contains
53\%, 61\%, 64\%,
and 65\% of source counts for the first through the fourth $\lambda$-bins,
respectively).

For each $\lambda$-bin, we calculate the average spectral flux
defined as
\begin{equation}
 \langle F_{\lambda}\rangle_i\equiv
\frac{\int_{\Delta\lambda_i} R_{\lambda}\lambda\ F_{\lambda}\,{\rm
d}\lambda} {\int_{\Delta\lambda_i} R_{\lambda}\lambda\, {\rm
d}\lambda}\equiv\frac{C_i}
{\int_{\Delta\lambda_i}
 R_{\lambda}\lambda\, {\rm d}\lambda}\, ,
\end{equation}
where
$C_i$ is the source count rate in the $i$-th
 $\lambda$-bin, and
$R_{\lambda}$
is the
system response (which includes the Optical Telescope Assembly
throughput and accounts for the grating and slit losses
and the finite size of the source extraction aperture; see \S
3.4.12 of the HST Data Handbook for STIS\footnote[6]{
http://www.stsci.edu/hst/stis/documents/handbooks/currentDHB/STIS\_longdhbTOC.html}
for details). The resulting flux values are given in
Table 1, while the spectrum is shown in Figure 2.
The total flux in the 1150--1700 \AA\ range $(\Delta\lambda=550$ \AA),
 can be estimated as
$F\simeq \Delta\lambda\,
\left(\sum_i \langle F_\lambda\rangle_i \Delta\lambda_i\right)
\left(\sum_i \Delta\lambda_i\right)^{-1} \simeq
(2.0\pm 0.2) \times 10^{-15}$ erg s$^{-1}$ cm$^{-2}$,
corresponding to the luminosity
$L_{FUV}=4\pi d^2 F =(4.7\pm 0.5)\times 10^{27} d_{140}^2$ erg s$^{-1}$.

In the original spectrum we see a brightening around
1372 \AA\ (marked with a vertical arrow in Fig.\ 1), resembling an
emission line. The count rate spectrum of the possible line
and its vicinity, binned to 1.16 \AA\
(4 high-resolution pixels) bins, is shown in Figure 3. The count excess,
$30\pm 10$ counts in the three bins, corresponds to the flux
$(4.0\pm 1.3)\times 10^{-17}$ erg cm$^{-2}$ s$^{-1}$ in the
1370.1--1373.6 \AA\ range. This spectral feature,
detected at a 3 $\sigma$ level, is clearly not
of a geocoronal origin (the nearest geocoronal oxygen line,
at 1356 \AA, is not seen in this observation), but we cannot rule out
the opportunity that it may be associated with an anomaly in the FUV-MAMA
background seen right above the feature (in the cross-dispersion direction).
Therefore, we consider this
as a marginal detection until the feature is confirmed in another
observation.

For the timing analysis, we used the
so-called TIME-TAG data files that contain the photon arrival times,
recorded at a $125$ $\mu$s time resolution, and high-resolution
detector coordinates (see \S2) associated with each of the events.
We extracted 2964 events from the above-defined four $\lambda$-bins,
with the height of extraction box equal to 9 high-resolution pixels
($\approx 22.5$\% of these counts are
expected to come from the source).
Note that the background count
rate in these data is about 3\% higher than that in the data used
for the spectral analysis where ``outstanding pixels'' were
filtered out (see above).
 The arrival times of the events
were corrected for the Earth and spacecraft motions
and transformed to
barycentric dynamical times (TDB) at the solar system barycenter,
using the STSDAS task ``odelaytime''.
 To correct the TDB arrival times for the effect of
the binary orbital Doppler shift (Taylor \& Weisberg 1989) and to
extract the light curve, we used the pulsar and binary orbit
ephemerides from van Straten et al.\ (2001).

Since the frequency of radio pulsations,
 $f=173.687948857032$ Hz
at the epoch of the STIS observation,
is known with high precision, much higher than one can achieve in the
time span $T_{\rm span}=24\,638$ s of our observation,
there is no need to search for the period in the FUV data.
We folded the corrected arrival times with the pulsar's period
and obtained the light curves for different numbers of phase bins.
An example of the light curve for 4 phase bins, with an arbitrarily fixed
 reference
phase (histogram) and averaged over the reference phase (smooth line; see
Gregory \& Loredo 1992 for a description of this procedure)
is shown in Figure 4.
The observed pulsed fraction
can be estimated as $f_{\rm p} = 4^{+5}_{-4}\%$, which means that the
pulsations are not statistically significant.

The upper limit on the pulsed fraction can also be estimated from
the $Z_n^2$ test (e.g., Buccheri et al.\ 1983) that gives $Z_n^2 =
2.7$, 2.8, 5.2, 8.1, and 8.7 for $n=1$, 2, 3, 4, and 5,
respectively, at the radio pulsation frequency ($n$ is the number
of harmonics included).
For randomly distributed arrival times, the probabilities of getting
$Z_n^2$ larger than these
values
are 0.26, 0.59, 0.52, 0.42, and 0.56,
respectively, which confirms that no periodic signal is
detected.
An upper limit on the pulsed fraction can be estimated from the $Z_1^2$ value,
assuming sine-like pulsations. At a $3\sigma$ confidence level, the limit
is 11\%--12\%, depending on the method used for estimation
(e.g., Protheroe 1987; Vaughan et al.\ 1994; Brazier 1994). It
translates into
an upper limit $f_{\rm p}^{\rm int} < 48\%$--53\%
 for the intrinsic pulsed fraction of J0437,
corrected for the background contribution.
The upper limit is close to the pulsed fraction at $E\gtrsim 1$
keV, and it exceeds the pulsed fraction, $\sim 30\%$, observed in
soft X-rays (Zavlin \& Pavlov 1998). If the FUV radiation is
thermal, as suggested by the spectrum (see \S4),
we expect the pulsed fraction to be even lower than 30\% (because 
the FUV radiation is presumably emitted from a large fraction of the NS surface),
so the
limit is not truly restrictive. The reason for that is
the very high
detector background. We attempted to improve the S/N repeating
the analysis for the first four orbits, with the lower
detector background, but obtained
approximately the same upper limit.
\section{Discussion}
\subsection{Neutron star or white dwarf?}
First of all, we should understand whether the observed FUV
spectrum is emitted by the WD companion or by the NS. The results of the
optical photometry
(Danziger et al.\
1993; Bailyn 1993) yield $T_{\rm{eff}}\approx 3750$--4500 K for the WD.
The dashed line in Figure 2 shows a blackbody spectrum (for
$T=4000$ K, $R=0.025R_{\sun}$,
$d=140$ pc)
that crudely fits the photometric data 
and clearly falls well below the FUV data points. More accurate
comparison should use cool WD atmosphere spectra (e.g., Bergeron
et al.\ 1995; Hansen 1999), which can be harder than the blackbody
at UV frequencies. We have applied a number of such models
(provided by P.\ Bergeron 2003, private communication) for various
chemical compositions, effective temperatures and gravitational
accelerations, scaling them to fit the optical data, and found
that the observed FUV flux exceeds the WD atmosphere model
predictions by several orders of magnitude in all the cases. Thus,
we conclude that the observed FUV emission originates from the NS.
\subsection{Thermal origin of the FUV spectrum}
We fit the spectrum with the absorbed power-law model,
$F_{\lambda}= F_{1500}\,(\lambda/1500\, {\rm
\AA})^{\alpha_{\lambda}}\times 10^{-0.4 A(\lambda)\, E(B-V)}$,
where the ultraviolet extinction curve $A(\lambda)$ is adopted
from Seaton (1979). The (low) interstellar extinction in the direction
of J0437 is poorly known; different authors use the color excess
$E(B-V)$ values from $<0.01$ (Bell et al.\ 1993) to 0.07
(Danziger et al.\ 1993). We performed the fits for three
values, $E(B-V)=0$, 0.03, and 0.07, and found the
power-law indices $\alpha_\lambda = -4.0\pm 1.2$, $-4.2\pm 1.2$,
and $-4.4\pm 1.2$, and the normalizations $F_{1500} = 2.6\pm 0.4$,
$3.4\pm0.5 $, and $4.4\pm 0.7 \times 10^{-18}$ erg cm$^{-2}$
s$^{-1}$ \AA$^{-1}$, respectively (see Fig.\ 5).
The inferred slope $\alpha_\lambda$
(albeit rather uncertain because of the strong
detector background)
is close to that of
the Rayleigh-Jeans
spectrum, $F_\lambda\propto \lambda^{-4}$. It indicates that, most
likely, the observed radiation
is {\em thermal radiation from
the NS surface} rather than magnetospheric radiation.
At $d=140$ pc, the normalization
parameter of the Rayleigh-Jeans fit corresponds to the brightness
temperature $T_5=0.215\, F_{1500} R_{13}^{-2}$ K,
where $T_5=T/(10^5\,{\rm K})$, $R_{13}$ is the radius of the emitting sphere
in units of 13 km.
(Here and below, the temperature and the
radius are given as measured by a distant observer.) Our fits
(with fixed $\alpha_\lambda=-4$) give $T_5 R_{13}^2 = 0.62\pm 0.06$, $0.79\pm
0.08$, and $1.1\pm 0.1$, for $E(B-V)=0$, 0.03, and 0.07,
respectively.

At $T\lesssim 10^5$ K the blackbody spectrum deviates appreciably from
the Rayleigh-Jeans limit in the FUV range. To investigate
the range of lower temperatures, we
also fit the absorbed blackbody  model to the observed
spectrum. The confidence contours in the $T$-$R$ plane
(Fig.\ 6) show that
 the lower limit on the surface temperature
is $4.2\times 10^4$ K, corresponding to a radius of 37 km, at a 68\% confidence
level ($3.1\times 10^4$ K and 59 km at a 90\% level).
 For a typical
 NS radius $R=13$ km, the inferred surface temperature is
$0.107\pm 0.010$, $0.121\pm 0.012$, and $0.156\pm 0.014$ MK, for
$E(B-V)=0$, 0.03, and 0.07, respectively.
The corresponding
bolometric luminosity can be estimated as
 $L_{\rm bol}=1.2\times 10^{29}\, T_5^4 R_{13}^2$
erg s$^{-1}$; for instance, $L_{\rm bol}=(2.6\pm 1.0)\times
10^{29}$ erg s$^{-1}$ for $E(B-V)=0.03$, $R=13$ km.

Because of the large uncertainties of the FUV data points,
deviations of the source spectrum from the Rayleigh-Jeans limit
cannot be established from the FUV data alone. Therefore, to infer
the upper limit on $T$ (lower limit on $R$),
we have to use the X-ray data. For the single-temperature blackbody
model, we find that $T$ should be lower than $0.23$ MK ($R>8.0$
km), for $E(B-V)=0.03$,
in order
not to exceed
the observed soft X-ray spectral flux (see
Fig.\ 7). It is clear from Figure 7 that the FUV and X-ray data
cannot be described by a single-temperature model.
Based on the X-ray observations,
Zavlin et al.\ (2002) suggested
a two-temperature model:
$T_{\rm core} = 2.1$ MK, $R_{\rm core}=0.12$ km; $T_{\rm rim}=0.54$ MK,
$R_{\rm rim}=2.0$ km, where the ``core'' and the ``rim'' correspond to
central and peripheral parts of the pulsar's polar caps.
We see from Figure 7 that extension of this model to the FUV range
underestimates the observed FUV spectral flux.
Most probably, this means that 
the NS surface temperature
is a function of the polar angle, such that it decreases from
$\sim 2$ MK at the magnetic poles to $\sim 0.1$ MK at the bulk
of the surface. Of course, the estimates for the brightness temperature
above do not imply a uniform surface temperature away from the
polar caps.
To determine the dependence of the surface temperature on magnetic
colatitude, phase-resolved spectroscopy is required.
\subsection { Internal heating}
The inferred surface temperature, $T\sim 10^5$ K,
is much higher than the temperatures $\lesssim 10^3$ K expected for
a passively cooling few-Gyr-old neutron star (e.g., Tsuruta 1998;
Schaab et al.\ 1999 [S99 hereafter]). Therefore, it requires a heating mechanism
to operate in J0437. If the sources of heat are in the highly
conductive (hence almost isothermal) NS interiors,
the surface temperature of 0.1 MK
implies the interior temperature of order 1~MK
(Gudmundsson, Pethick, \& Epstein 1983).
Various mechanisms of internal NS heating were investigated
in a number of
works (e.g., S99,
and reference therein).
The mechanisms relevant to MSP heating
can be divided into two groups.
Firstly, heat can be produced by the dissipation of energy of differential rotation
caused by frictional
interaction between the faster rotating superfluid core and the slower
rotating outer solid crust (Shibazaki \& Lamb 1989; Larson \& Link 1999,
and references therein).
The heating mechanisms from the second group are associated with 
readjustment of the NS structure to a new equilibrium state as star
rotation slows down. If the NS crust is solid, it will undergo
cracking when the tension force exceeds a critical value, and the crust
strain energy will be released as heat (Cheng et al.\ 1992).  In
addition, as the star spins down, the centrifugal force decreases
and central density increases
causing the shift in
chemical equilibrium,
which modifies the rate of nuclear reactions
and may lead to heat release (Reisenegger 1995).

The high surface temperature of J0437
strongly constrains heating
mechanisms.
For instance, such a temperature rules out
crust cracking and chemical heating of the crust as the main heat
sources (cf.\ Fig.\ 7 of S99).
Chemical heating of the core and frictional heating
remain viable mechanisms. For the latter, one can constrain the excess
angular momentum $\Delta J_s$, residing in the superfluid, and the
angular velocity lag $\bar{\omega}$, between the superfluid and the
crust, averaged over superfluid moment of inertia. Using equation (9)
of Larson \& Link (1999), we obtain $\Delta J_s = L_{\rm bol}/|\dot{\Omega}|
=1.1\times 10^{41}\, T_5^4 R_{13}^2$ erg s, $\bar{\omega}=\Delta J_s/I_s
=0.15\, T_5^4 R_{13}^2$ rad s$^{-1}$,
where $\dot{\Omega} = -1.086\times 10^{-14}$ rad s$^{-2}$
is the time derivative of the angular frequency of the pulsar,
$I_s=7.3\times 10^{43}$ g cm$^2$ is the moment of inertia
of the differentially rotating portion of the superfluid, estimated
for the Friedman \& Pandharipande (1981) equation of state.
Because other mechanisms can contribute to heating of J0437,
these estimates of $\Delta J_s$ and $\bar{\omega}$ should be considered
as upper limits.
\subsection{
External heating}
In addition to the above-discussed internal sources,
heat can be provided by
relativistic particles
created in the pulsar's
acceleration zones and
bombarding the NS surface
(e.g., Harding \& Muslimov 2002, and references therein).
The energy
of these particles is released in polar caps at the NS
magnetic poles,
with a radius $r_{\rm{pc}}\sim (R^{3}\Omega/c)^{1/2}\approx 2\,
(R/10~\rm{km})^{3/2}(P/6~\rm{ms})^{-1/2}$ km.
Moreover, inward-directed radiation from this particle population or
even precipitating secondary particles created in the closed
field line zone can heat the surface outside
of the traditional polar caps (Wang et al.\ 1998).
The energy dissipated in the
NS polar regions per unit time
(heating luminosity) depends on the geometry
of the acceleration region
and the degree of electric field screening. The numerical calculations, carried out
by Harding \& Muslimov (2002) for the case of MSPs, give heating
rates somewhat lower than (but close to) the
bolometric
luminosity inferred from the X-ray observations of
 J0437, $L_{\rm{bol}}^{\rm pc}\approx 2\times 10^{30}$ erg s$^{-1}$
(Zavlin et al.\ 2002).
A major fraction of the thermal energy
deposited by the relativistic particles propagates
towards the surface and is radiated in soft X-rays from the
polar cap(s). However, some
energy flows inwards,
heating deeper NS layers
beneath a larger surface area.
If a fraction of this energy
reaches the inner crust, it spreads over the whole interior of
the NS because of its very high thermal conductivity, so
eventually it will be radiated from the entire NS surface, together with the
heat supplied by possible internal heating mechanisms.
The effect of the external heating on the temperature distribution
in an NS requires numerical solution of a complicated problem
of heat transport, with allowance for radiation from the surface.
This problem has not been solved, to the best of our knowledge,
so it is hard to assess the contribution of the polar cap heating
to the observed FUV radiation. We can only state that if the external
heating dominates, then a fraction
$\sim L_{\rm bol}^{\rm surf}/L_{\rm bol}^{\rm pc}\sim 0.1$--0.5 of the
energy deposited at the polar cap goes to heating of the NS. Crude estimates
show, however, that such a large fraction of heat flowing inwards
 would require thermal conductivities much higher
than calculated for the NS crust (e.g., Jones 1978), suggesting that
particle precipitation only on the open field line zone cannot account
for general surface heating of old NSs.

In principle, a very old NS could be heated by exotic particles
(or products of their decay)
captured in the NS matter. For instance, Hannestad, Ker\"anen, \& Sannino
(2002) consider NS heating by products of decay of
Kaluza-Klein majorons and gravitions, that form a halo around the NS,
and conclude that the thermal luminosity of an old NS should reach a
constant level corresponding to the energy deposited by decay particles.
(See Pavlov, Stringfellow, \& C\'{o}rdova  1996 for references to earlier papers.)
Although
the Occam razor 
suggests that less exotic mechanisms should be explored first,
the possibility to indirectly investigate exotic particles through
thermal radiation from MSPs cannot
be dismissed at the present stage.
\subsection{
Are millisecond pulsars hotter than ordinary old neutron stars?}
It is interesting to compare the inferred surface temperature of J0437
with the temperatures (or upper limits) for other old pulsars, including MSPs.
We compiled the results of the available UV-optical
 observations of old ($\gtrsim 1$ Myr),
nearby ($\lesssim 300$ pc) NSs in Table 2 and re-estimated the temperatures
using the most recent results on the distances.
In addition to the old pulsars, we included the famous
``dim'' isolated NS, RX~J1856.5$-$3754, whose age is unknown, because it might
be an MSP (Pavlov \& Zavlin 2003).
We see from Table 2 that the temperature of J0437 is the lowest {\em measured}
temperature for a NS. The observations of the other two
nearby MSPs, J2124$-$3358
and J0030+0451, were not nearly as deep as required to detect their thermal
radiation at a similar temperature. All the upper limits on $T$ in Table 2
are also above the temperature of J0437 with {\em one exception} --- $T_5 < 0.88$
for PSR J0108$-$1431, whose spin-down age is a factor of 30 smaller
than that of J0437\footnote{Although the true age of a pulsar 
may differ substantially from its 
spin-down age, the low temperature of the WD companion implies
(e.g., Hansen \& Phinney 1998) that J0437
is older than 3 Gyr, that is a factor of 18 larger than the spin-down age
of PSR J0108$-$1431.}.  
Although the difference in the temperatures is not very
large, it should be noted that the upper limit in Table 2
(see also Mignani et al.\ 2003b)
is rather conservative, in the sense that it is obtained assuming $d=200$ pc,
the largest among the distances, 60--200 pc, estimated from various models
of Galactic electron distribution (Tauris et al.\ 1994; Taylor \& Cordes 1993;
Cordes \& Lazio 2002). For instance, the limit is $T_5 < 0.45$ for $d=130$ pc
(Mignani et al.\ 2003a). Even if the true distance is somewhat larger than
200 pc, the upper limit remains surprisingly low, given the much younger
spin-down age of J0108$-$1431.

The higher temperature of the older J0437 cannot be attributed to accretion
in the binary system because 
the accretion ceased about 3 Gyrs ago, while the time for a NS to cool to
$10^5$ K, losing the accretion-generated heat, is $\lesssim 10$ Myr.
The high temperature can possibly be explained by
stronger frictional heating, consistent with a factor of 15 larger $|\dot{\Omega}|$.
It is also possible that ``rotochemical heating'' of the NS core 
(Reisenegger 1995,1997)
 has been turned on in this MSP, or J0437 is simply sufficiently old to
accumulate enough heat by capturing 
exotic particles (see \S4.4). Finally, the higher temperature of J0437
might be associated with stronger magnetospheric heating, since its $\dot{E}$
is about 700 times larger than that of J0108$-$1431. In this case
one should expect MSPs with higher $\dot{E}$ (up to 3 orders of magnitude
in most powerful MSPs, such as PSR B1821$-$24) to be even hotter than J0437,
but it would be difficult to discern their surface emission
since these pulsars are also strong sources of magnetospheric radiation.

Whatever is the reason of the higher temperature of J0437, it is tempting
to assume 
that the surface temperatures of MSPs are generally higher than those
of old ``ordinary'' pulsars of similar or even younger ages.
Since ordinary isolated NSs cool down with age even in the presence of 
internal heating
(Tsuruta 1998; S99), this assumption implies that
the transition from ordinary pulsars to MSPs is accompanied by
a considerable growth of NS temperature.
Moreover, the temperature is maintained high in the course of 
the Gyr-long MSP thermal evolution by a heating process, probably
associated with specific properties of MSPs, such as fast rotation or 
old age. To understand the apparently ``nonmonotonous'' thermal evolution 
of old NSs, thermal emission from a larger sample of these objects should
be investigated.
\subsection{Possible spectral line at 1372 \AA}
Although the detection of the emission line at 1372 \AA\
($h\nu = 9.04$ eV) is marginal,
it is interesting to speculate
as to
about its origin.
Firstly, this could be an electron cyclotron line
(e.g., formed in a corona above the polar cap) in the magnetic
field $B=7.8\times 10^8 (1+z)$ G, where $z$ is the gravitational redshift.
Such a field is
close to the ``conventional'' magnetic field,
$B_p =
6.6\times 10^8\, R_6^{-3} I_{45}^{1/2}
(\sin\alpha)^{-1}$ G,
at the magnetic
poles of an NS losing energy via magnetodipole radiation
($R=10^6 R_6$ cm and $I=10^{45}I_{45}$ g cm$^2$ are the NS radius and moment
of inertia, $\alpha$ is the angle between the magnetic and rotational axes).
The main
difficulty with this interpretation is the small width of the
observed line, $\sim 3$ \AA. First, it implies a very uniform field
in (hence, a small size of) the emitting region,
$\Delta B/B < \Delta\lambda/\lambda$. For instance, if the
emitting region is a hot spot at the magnetic pole, its size should be
smaller than $\sim 0.5$ km.
Second, the thermal (Doppler)
width of the cyclotron line is $\Delta\lambda_D = \lambda (2kT/m_ec^2)^{1/2}|\cos\theta|
\approx 25\, T_6^{1/2} |\cos\theta|$ \AA, where $\theta$ is the angle between the
line of sight and the direction of the magnetic field.
For the thermal width to be smaller than $\Delta\lambda$,
a phase-averaged $|\cos\theta|$ must be
$\lesssim 0.12\, T_6^{-1/2}$ ($\theta \gtrsim 83^\circ$
for $T_6=1$), which is hard to reconcile with the constraints
on the rotation and magnetic inclinations obtained from the radio
polarimetry (e.g., Manchester \& Johnston
1995).

Alternatively, if the NS surface is covered with
a hydrogen atmosphere with overheated outer layers (e.g., due to
convection), the observed line may be one of the three
gravitationally redshifted Zeeman
components of the Ly${\alpha}$ line in a magnetic field $B\sim
10^8$--$10^9$ G.
For instance, at $B=7\times 10^8$ G, the wavelengths of the Zeeman
components at the NS surface are 730, 1069, and 1334 \AA\ (Ruder
et al.\ 1994). If the observed line is the redshifted
 $\pi$-component, the redshift is $z=0.28$,
so the wavelengths of the redshifted $\sigma$-components, 937 and
1712 \AA, are outside of the observed FUV range.
 The thermal width
of the Zeeman component, $\sim 0.6\,T_6^{1/2}$ \AA, is much
smaller than the observed width. The line broadening can be caused
by a magnetic field nonuniformity: $\Delta\lambda =3$ \AA\
corresponds to $\Delta B/B \sim 0.3$ (estimated from Fig.\ A.2.1
of Ruder et al.\ 1994), that is the putative corona may spread
over a substantial fraction of the NS surface. The luminosity in
the line, $L_{1372}\approx 0.9\times 10^{26}$ erg s$^{-1}$, is a
small fraction, $\approx 2\%$, of the observed FUV luminosity.
If the emitting region is in collisional (coronal) equilibrium,
the corresponding emission measure is $n_{H}^2 V \sim 10^{49}$--$10^{50}$
cm$^{-3}$ for $T\sim 0.1$--1 MK.
If the 1372 \AA\ line is confirmed in a future FUV observation,
the most convincing confirmation of the Ly$\alpha$ interpretaion
would be detection of another Zeeman component 
(e.g., the $\sigma$-component in the NUV range), which would allow
one to measure both the magnetic field and the gravitational redshift.
\subsection {Magnetospheric component}
 Zavlin et al.\ (2002) found that, in
addition to the two-temperature thermal component, a power-law component with
a photon index $\Gamma=2.2^{+0.3}_{-0.6}$ is needed to fit the X-ray
spectrum at energies above 2 keV (Fig.\ 7). However, extension of this
component with the best-fit $\Gamma$ to the FUV range is well above
the observed FUV spectrum, and
even for $\Gamma=1.6$ the extension is only marginally
consistent with the FUV data (Fig.\ 7).
We can crudely estimate the upper limit on the magnetospheric luminosity
is the FUV range as $L_{\rm FUV}^{\rm pl} \lesssim 4.2\times 10^{27}$ erg s$^{-1}$.
The apparent discrepancy can be explained assuming
that the spectrum of the magnetospheric radiation
breaks down when the frequency decreases from X-rays to the FUV range.
The optical upper limits for MSPs J0030+0451 (Koptsevich et al.\
2003) and J2124$-$3358 (Mignani \& Becker 2003), well below the
extensions of their X-ray power-law components, suggest that it might
be a common property of MSPs,
contrary to, e.g, ordinary middle-aged pulsars.
Alternatively, as Zavlin et al.\ (2002)
noticed, it is possible that the excess counts
at higher energies in MSP spectra
 might be interpreted as thermal radiation from a very hot, small area
within the polar cap
(e.g., $T\sim 12$--15 MK, $R\sim 1$--2 m for J0437).
To understand which of the two interpretations is correct, deeper observations
of J0437 in hard X-rays are required.
\section{Summary and conclusions}
The STIS/FUV-MAMA observation of J0437 provided first firm
detection of an MSP in the optical-UV range. The FUV spectrum is
best interpreted as thermal emission from the NS surface with a
temperature of about 0.1 MK. This temperature exceeds the upper
limit on the temperature of the younger, but less luminous, ordinary pulsar
J0108--1431. This is likely associated with a difference in
spindown-driven heating. If magnetospheric heating plays a role,
it must be effectively communicated, perhaps by radiation or
secondary particles, to the bulk of the NS surface. Evolutionary
differences between ordinary pulsars and MSPs might plausibly affect
the internal thermal history.
To understand thermal evolution of old NSs,
more MSPs and ordinary old pulsars should be observed in the
optical-UV range.

Comparison of the FUV and X-ray spectra shows that the temperature
is not uniformly distributed over the NS surface. The X-ray
observations, sensitive to higher temperatures, show a smaller
size of the emitting region, naturally interpreted as a pulsar
polar cap,
perhaps also with a nonuniform temperature. To understand the
temperature distribution over the NS surface, phase-resolved
spectroscopy in both X-rays and FUV is needed. We failed to detect
FUV pulsations because the source was placed at a region of high
detector background. An FUV observation of J0437 with an optimal
positioning on the detector could detect pulsations (or put a
stringent limit on the pulsed fraction) and provide information on
the temperature distribution.

The FUV upper limit on the nonthermal (magnetospheric) component,
observed in hard X-rays, suggests a spectral turnover of this
component at EUV wavelengths, which can be a generic property of
MSPs. However, the upper limit is not very strong because of
large errors associated with the high detector background. To
tightly constrain the nonthermal component, another FUV-MAMA
observation with improved S/N as well as deep NUV and X-ray
exposures are required.

The marginally detected emission line at 1372 \AA\ can be
interpreted as an electron cyclotron line or, more likely, a
Zeeman component of the Hydrogen Ly$\alpha$ line in a magnetic field of $\sim 10^9$
G. Confirming this line would be of profound importance as it
provides an opportunity to directly measure the MSP magnetic field
and gravitational redshift (if another Zeeman component is
also detected).

\acknowledgements

 We are grateful to Kailash Sahu and Mike Potter of STScI for the advice on the data
 analysis. We thank
Pierre Bergeron and Brad Hansen for providing the
 white dwarf atmosphere models, 
Slava Zavlin for the fits of the X-ray spectra,
and Jules Halpern for the help in preparing
the proposal for this program.
The fruitful discussions with Dima Yakovlev, Mal Ruderman, Divas Sanwal,
and Andreas Reisenegger
are acknowledged. 
Support for program GO-9098 was provided by NASA
through a grant from the Space Telescope Science
Institute, which is operated by the Association of
Universities for Research in Astronomy, Inc., under NASA
contract NAS 5-26555.
 This work was also partly supported by NASA grant NAG5-10865.

\newpage
\begin{figure}[H]
 \centering
\includegraphics[width=5.0in,angle=0]{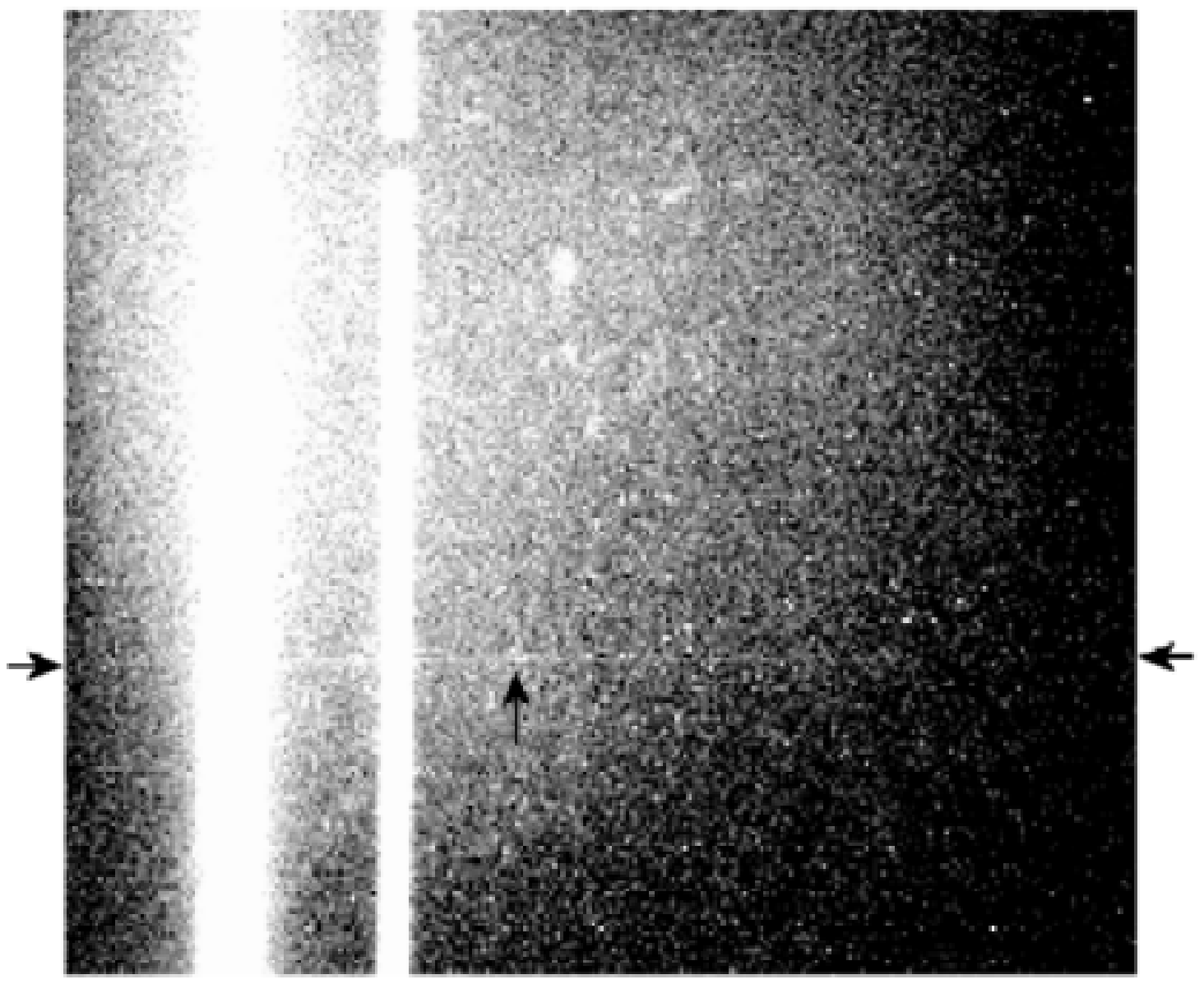}
\caption{Distribution of counts on the FUV-MAMA detector.
The spatial (Y)
and dispersion (X) axes
are in the vertical and horizontal
  directions, respectively. The spectrum of J0437 is shown by horizontal arrows. The background
 is clearly dominated by the nonuniformly distributed ``thermal
 glow'' which is the strongest at the upper
 left corner and the weakest at the bottom.
The vertical arrow shows the location of the possible spectral feature
  (see \S4.6  and Fig.\ 3).}
\end{figure}

\begin{figure}[H]
 \centering
\includegraphics[width=5.0in,angle=90]{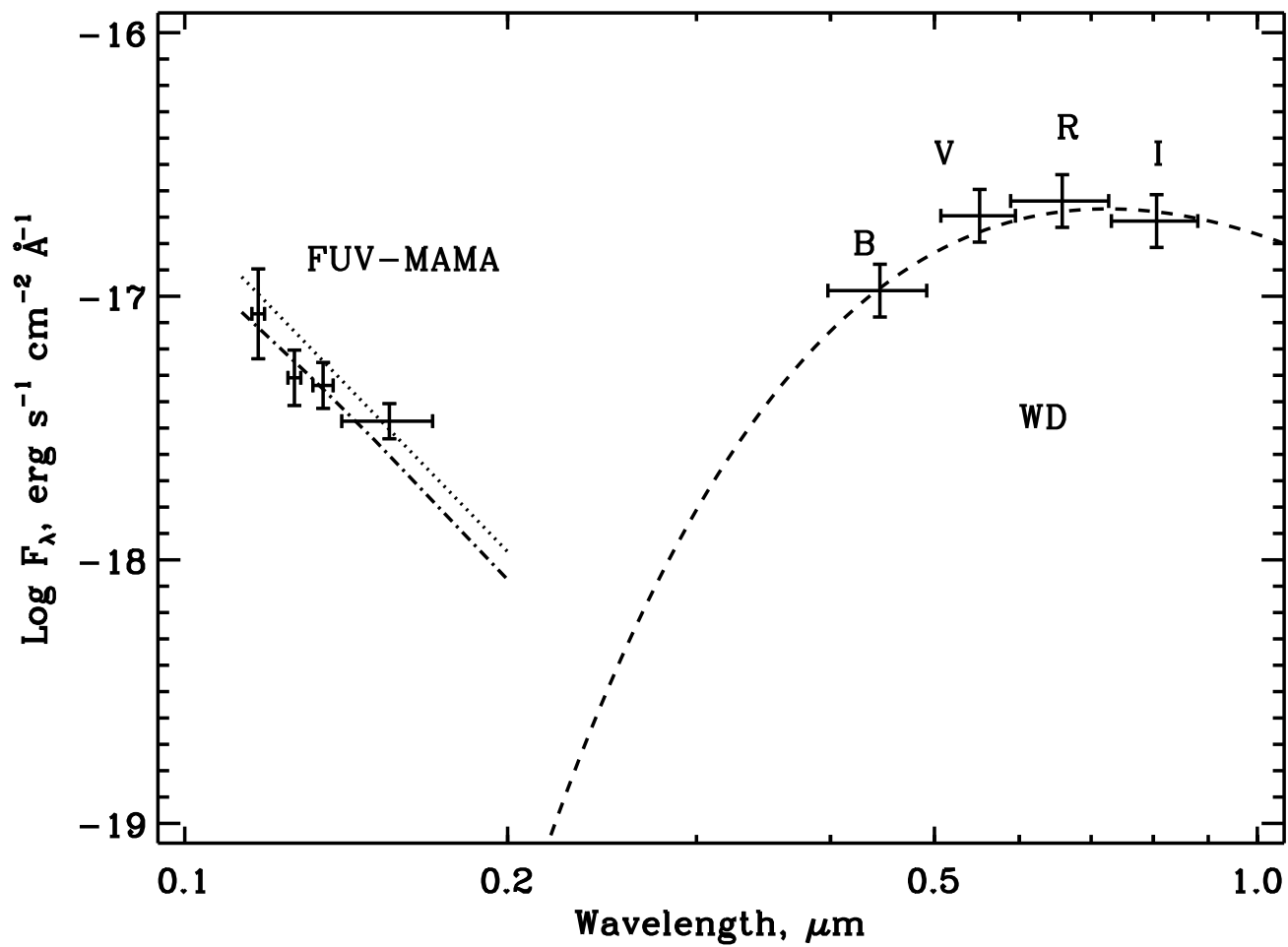}
\caption{FUV spectrum of J0437. The error bars on the left
represent the measured average fluxes in four $\lambda$-bins. 
The dash-dotted line is the fit
with the absorbed power-law model for $E(B-V)=0.03$. The dotted line
shows the same model but dereddened. The dashed line
is a blackbody spectrum with the temperature of 4\,000 K,
fitting the
dereddened B,V,R,I WD fluxes, taken
from Danzinger et
al.\ (1993).}
\end{figure}

\begin{figure}[H]
 \centering
\includegraphics[width=4.5in,angle=90]{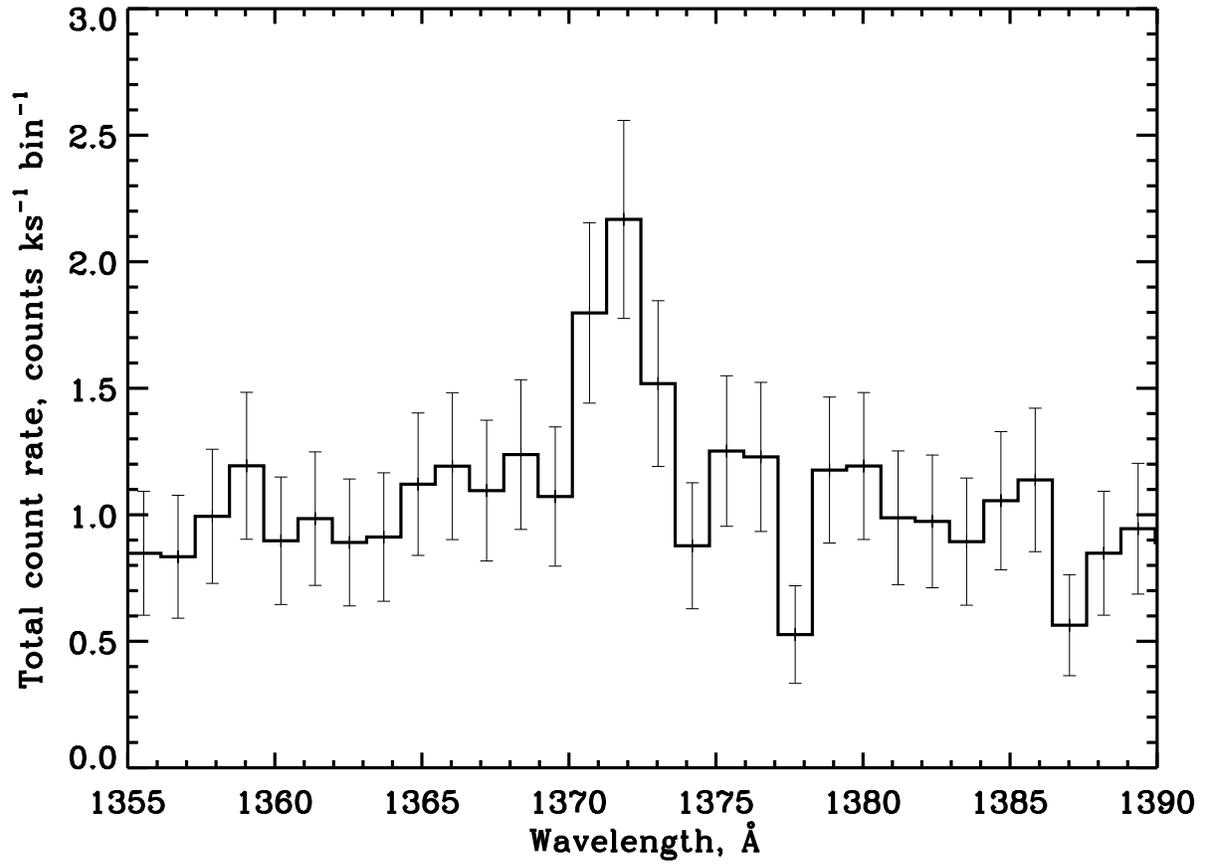}
\figcaption{Total (source + background) count rate distribution in the region around the
possible emission feature at 1372 \AA.}
\end{figure}

\begin{figure}[H]
 \centering
\includegraphics[width=4.5in,angle=90]{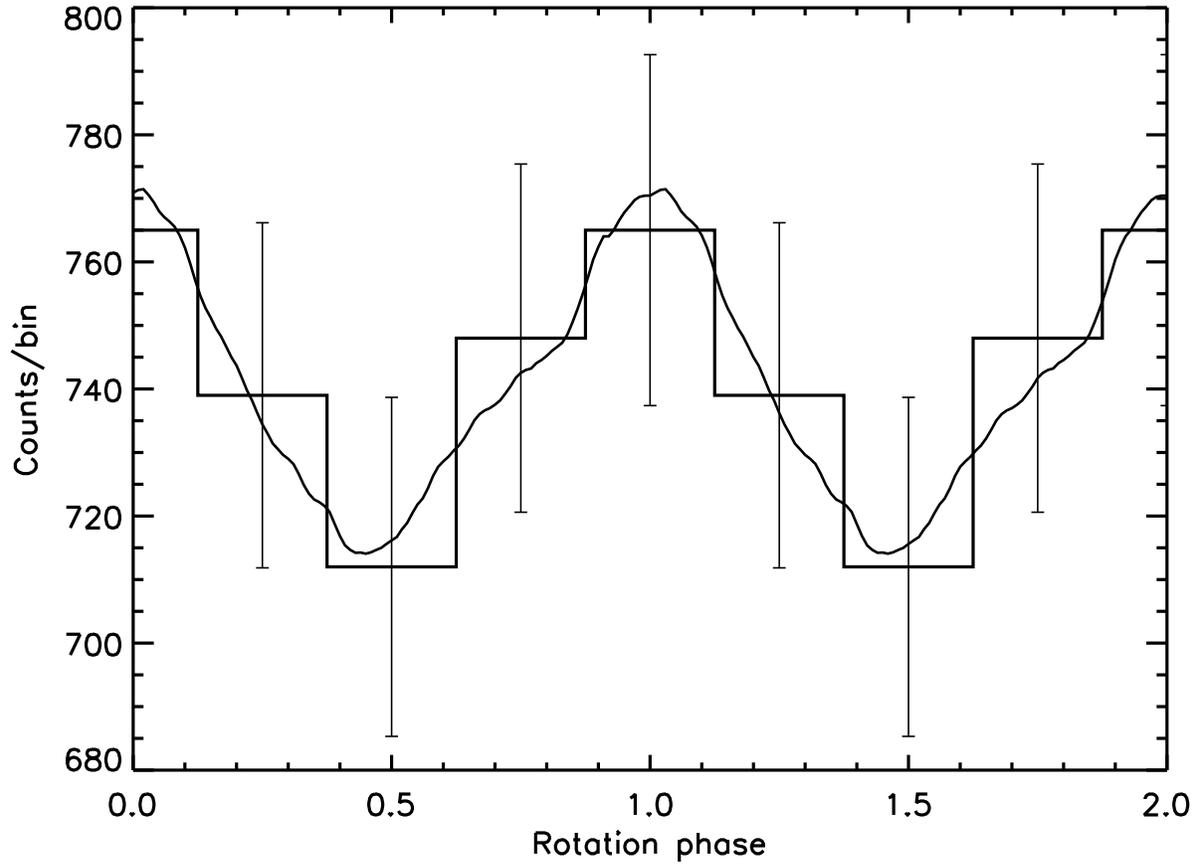}
\figcaption{
Pulsar light curve for 4 phase bins.
The smooth line shows the light curve averaged over the reference phase (see
\S2). }
\end{figure}

\begin{figure}[H]
 \centering
\includegraphics[width=5.0in,angle=90]{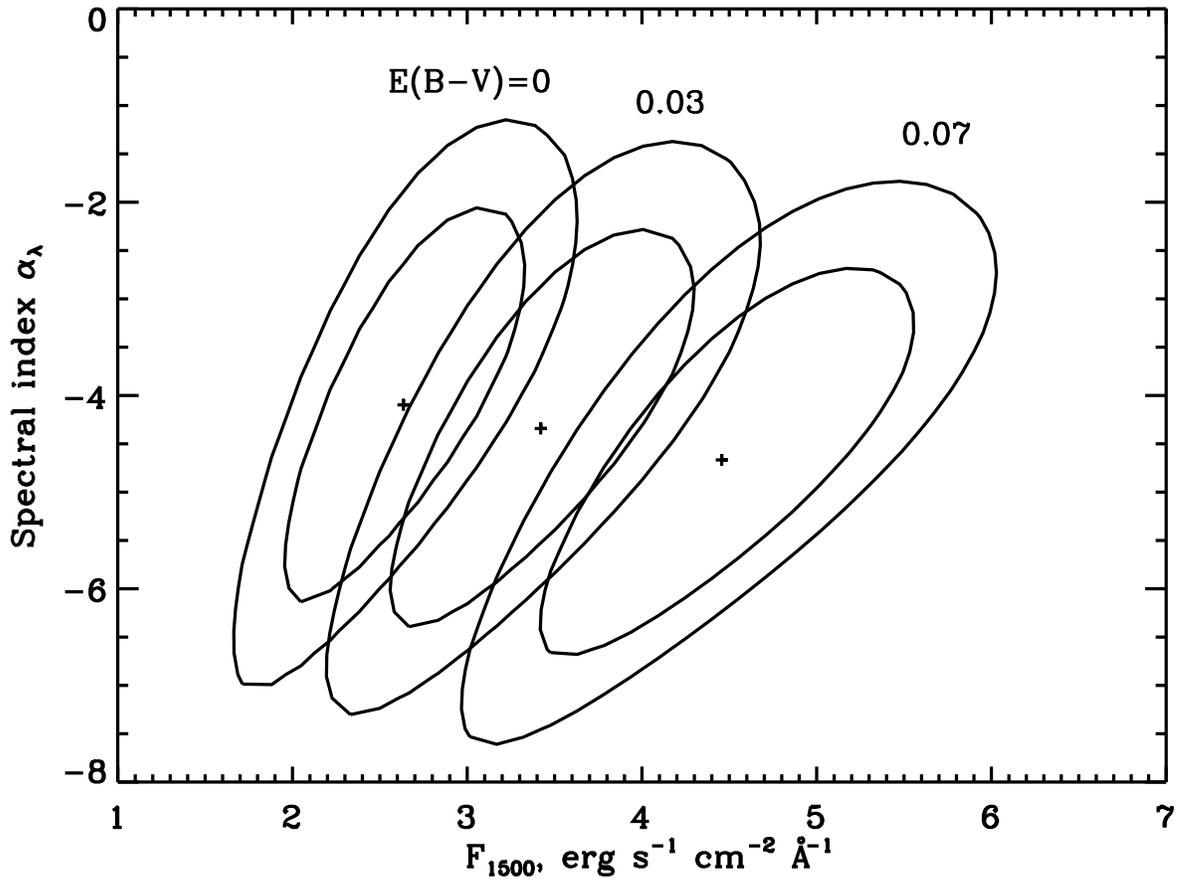}
\caption{ Confidence levels (68\% and 90\%)  for the absorbed
power-law model fit, for $E(B-V)=0$, 0.03, 0.07.}
\end{figure}

\begin{figure}[H]
 \centering
\includegraphics[width=4.5in,angle=90]{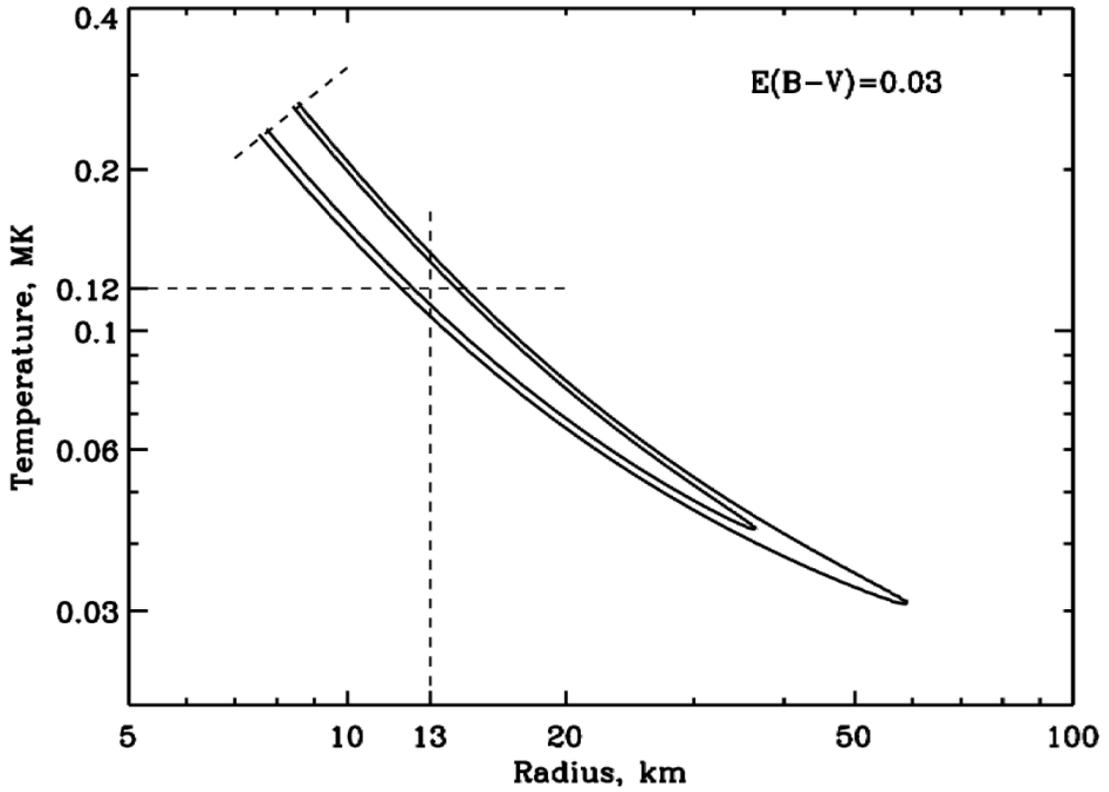}
\figcaption{Confidence levels (68\% and 90\%)  for the absorbed
blackbody model fit, for $E(B-V)=0.03$. The contours are cut 
(the dashed line in the top left corner) because the X-ray model flux exceeds
the observed one at higher temperatures and smaller radii (see Fig.\ 7).
The horizontal dashed line shows the temperature
corresponding to a plausible NS radius of 13 km.}
\end{figure}

\begin{figure}[H]
 \centering
\includegraphics[width=5.0in,angle=90]{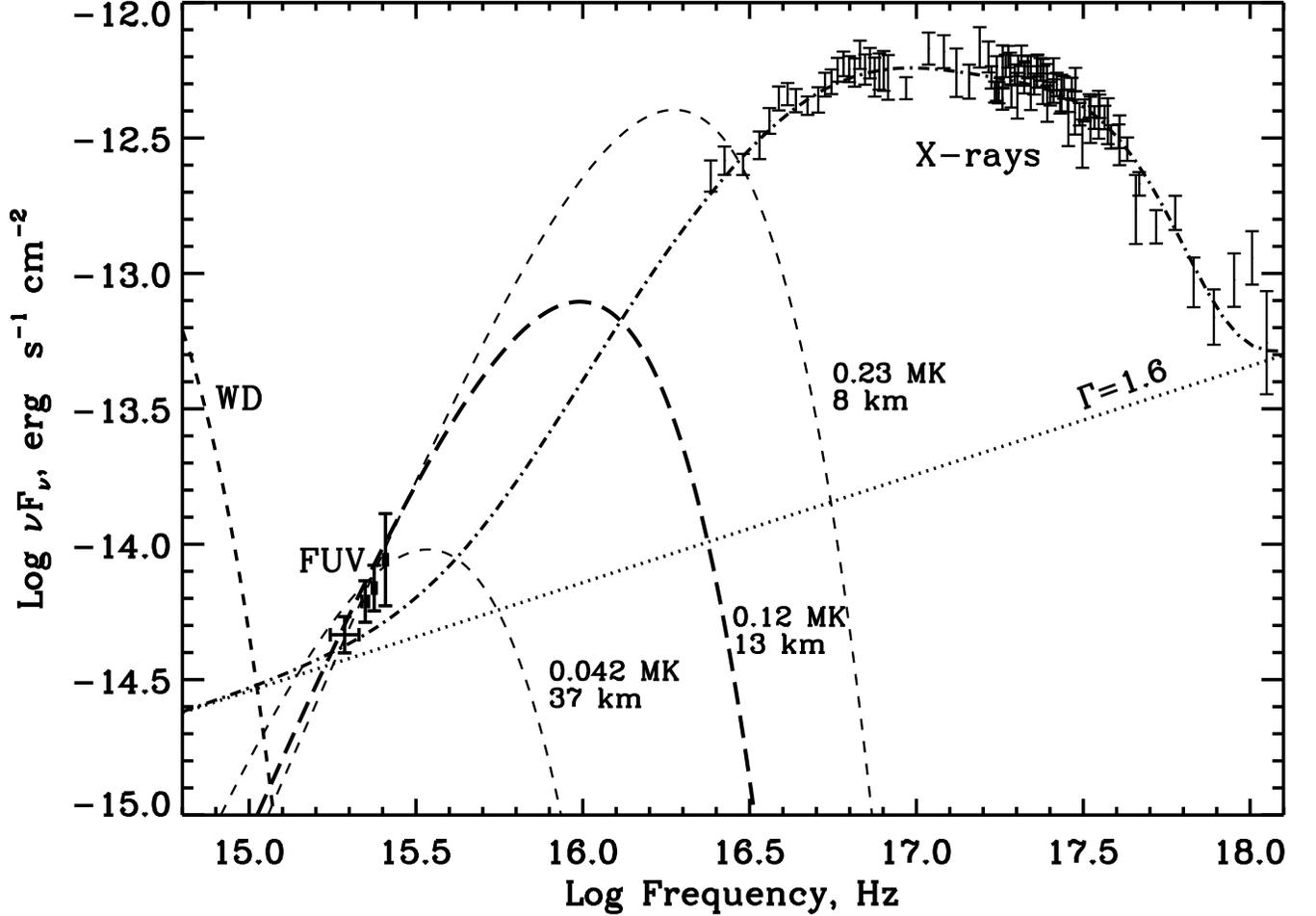}
\caption{Multiwavelength spectrum of J0437. The data points
on the
left represent the measured FUV fluxes; the unabsorbed
(dereddened) fluxes are only slightly higher for plausible
values of $E(B-V)$ (see Fig.\ 2). The
data points on the upper right
are the unabsorbed X-ray fluxes from {\sl Chandra} ACIS and {\sl
ROSAT} PSPC observations (Zavlin et al.\ 2002). Three dashed lines 
marked with temperature and radius values show the 
range of formally acceptable blackbody models (see \S4.2 for details).
The dashed line
marked ``WD'' corresponds to the blackbody with the temperature of 4\,000 K
(cf.\ Fig.\ 2).
The dash-dotted line that goes through the X-ray data points is the
two-temperature hydrogen atmosphere + power-law model (see \S4.2, \S4.7, and
Zavlin et al.\ 2002). The power-law component with the photon
index $\Gamma=1.6$ is shown separately by the dotted line.}
\end{figure}
\newpage
\begin{table}[htb]
\caption[]{Counts and fluxes in $\lambda$-bins}
\begin{center}
\begin{tabular}{ccccccccc}
\tableline\tableline
$\lambda$-bin & $N_{t}$ & $N_{b}$\tablenotemark{a} & $\delta N_{b}$ & $\delta N_{b}^{\rm uni
}$  & $N_{s}$ & $\delta N_{s}$ & S/N & $F_\lambda$\tablenotemark{b}\\
(\AA) & & & & & & & & \\
\tableline
1155-1187 & 367 & 313 & 9.3 & 6.2& 54.3 & 21.3& 2.5 & $7.5\pm2.6$ \\
1248-1283 & 615 & 465 & 15.0 & 7.6 & 149.3 & 28.0 & 5.1 & $5.4\pm0.9$ \\
1316-1376 & 816 & 609 & 22.6 & 8.7 & 206.7 & 36.5 & 5.8 & $4.6\pm0.6$ \\
1400-1702 & 1369& 1078 & 25.1& 11.6& 290.6& 44.7 & 6.5 & $2.9\pm0.4$ \\
Summed\tablenotemark{c}& 3167 & 2466 & 38.2& 17.5& 700.8& 68.0& 10.3 & $3.7\pm0.3$ \\
\tableline
\end{tabular}
\end{center}
\tablenotetext{a}{Number of background counts scaled to the 5 pixel extraction box.}
\tablenotetext{b}{Observed average spectral flux, in units of $10^{-18}$ erg s$^{-1}$ cm$^{-
2}$ \AA$^{-1}$, corrected for the finite aperture.}
\tablenotetext{c}{Values for summed $\lambda$-bins.}
\end{table}
\begin{table}[htb]
\caption[]{
        Blackbody temperatures of nearby
        old neutron stars
        }
\begin{center}
\begin{tabular}{lccccccc} \tableline \tableline
NS & $T$\tablenotemark{a} & Distance\tablenotemark{b} & Age\tablenotemark{c}   &
$F_\nu$\tablenotemark{d} & $\lambda$\tablenotemark{e} & $E(B-V)$\tablenotemark{f} & Ref.\tablenotemark{g} \\
     & ($10^5$ K)  &  (pc) & (Myr) & ($10^{-31}$ cgs) & ($\mu$m)  &  & \\
  \tableline
B1929+10          &      $<4.9$   & $331\pm 10$  & 3.2  & 5.4 & 0.24& 0.10 & 1 \\
B0950+08            &      $<2.5$   &  $262\pm 5$  &  16  & 5.1 & 0.31 & 0.01 & 2 \\
J0108$-$1431        &      $<0.88$   &  200  &   170  & $< 1.5$ & 0.49 & 0.01 & 3 \\
J1856$-$3754        &      5.6\tablenotemark{h} &  $117\pm 12$ &  ? &  240 & 0.15 & 0.03 & 4
 \\
J0437$-$4715   &      1.2 &   $139\pm3$ & 4900 & 20 & 0.15 & 0.03 & 5 \\
J2124$-$3358    &      $<4.6$    &   270  &  7200 & $<3.3$ &  0.49 & 0.05 & 6 \\
J0030+0451      &      $<9.2$   &    320  &  7700   & $<4.8$ & 0.49 & 0.05 & 7 \\
\hline
 \end{tabular}
\end{center}
\tablenotetext{a}{Blackbody (brightness) temperature from optical/UV observations,
 for an NS radius of 13 km and quoted distance and color index.}
\tablenotetext{b}{Distances estimated from parallaxes (with $\pm$
uncertainties; see Brisken et al.\ 2002; Walter \& Lattimer 2002;
van Straten et al.\ 2001) or from the dispersion measure
(Cordes \& Lazio 2002).}
\tablenotetext{c}{Characteristic age of pulsars, $\tau= P/(2\dot{P})$.}
\tablenotetext{d}{Observed spectral flux (or upper limit),
in units of $10^{-2}$ $\mu$Jy, at the wavelength quoted in the next
column.}
\tablenotetext{e}{Reference wavelength.}
\tablenotetext{f}{Adopted color index.}
\tablenotetext{g}{References for the spectral flux measurements.--(1) Mignani et
al.\ 2002; (2) Pavlov et al.\ 1996; (3) Mignani et al.\ 2003b; (4)
Pons et al.\ 2002;
(5) this work; (6) Mignani \& Becker 2003; (7) Koptsevich et al.\ 2003.}
\tablenotetext{h}{Lower temperatures, $T_5<3.9$, corresponding to $R>16$ km, are required in
a two-component blackbody model (Pavlov et al.\ 2002; Braje \& Romani 2002).}
\end{table}
\end{document}